\documentclass[journal]{IEEEtran}

\usepackage{amsmath} 
\usepackage{amssymb} 
\usepackage{graphicx} 
\usepackage{cite} 
\usepackage{subcaption} 
\usepackage{booktabs} 
\usepackage{float} 
\usepackage{xcolor} 
\usepackage{url} 

\newcommand{\vect}[1]{\mathbf{#1}}
\newcommand{\matr}[1]{\mathbf{#1}}

\newcommand{\ident}{\mathbf{I}}
\newcommand{\noisevar}{\sigma^2}
\newcommand{\SNR}{\text{SNR}}
\newcommand{\BER}{\text{BER}}
\newcommand{\EE}{\text{EE}}
\newcommand{\PAPR}{\text{PAPR}}
\newcommand{\CCDF}{\text{CCDF}}

\title{MD-OFDM: An Energy-Efficient and Low-PAPR MIMO-OFDM Variant for Resource-Constrained Applications}

\author{
    \IEEEauthorblockN{Rahul Gulia}
    \IEEEauthorblockA{Rochester Institute of Technology\\
    Rochester, NY, USA\\
    rg9828@rit.edu}
}

\begin{document}

\maketitle

\begin{abstract}
Orthogonal Frequency Division Multiplexing (OFDM) combined with Multiple-Input Multiple-Output (MIMO) techniques forms the backbone of modern wireless communication systems. While offering high spectral efficiency and robustness, conventional MIMO-OFDM, especially with complex equalizers like Minimum Mean Square Error (MMSE), suffers from high Peak-to-Average Power Ratio (PAPR) and significant power consumption due to multiple active Radio Frequency (RF) chains. This paper proposes and mathematically models an alternative system, termed Multi-Dimensional OFDM (MD-OFDM), which employs a per-subcarrier transmit antenna selection strategy. By activating only one transmit antenna for each subcarrier, MD-OFDM aims to reduce PAPR, lower power consumption, and improve Bit Error Rate (BER) performance. We provide detailed mathematical formulations for BER, Energy Efficiency (EE), and PAPR, and discuss the suitability of MD-OFDM for various applications, particularly in energy-constrained and cost-sensitive scenarios such as the Internet of Things (IoT) and Low-Power Wide Area Networks (LPWAN). Simulation results demonstrate that MD-OFDM achieves superior BER and significantly lower PAPR compared to MMSE MIMO, albeit with a trade-off in peak overall energy efficiency due to reduced spectral multiplexing.
\end{abstract}


\section{Introduction}
Orthogonal Frequency Division Multiplexing (OFDM) is a cornerstone technology in modern wireless communication systems due to its robustness against multipath fading and high spectral efficiency \cite{ETSI_OFDM}. When combined with Multiple-Input Multiple-Output (MIMO) techniques, MIMO-OFDM significantly boosts data rates and system capacity by enabling spatial multiplexing or diversity \cite{Foschini1998, Telatar1999}. However, conventional MIMO-OFDM systems, particularly those employing multiple active transmit antennas and complex linear equalization techniques like Minimum Mean Square Error (MMSE), face significant challenges: high Peak-to-Average Power Ratio (PAPR) \cite{Wang2005} and substantial power consumption due to multiple active Radio Frequency (RF) chains \cite{Bjornson2016}. High PAPR necessitates highly linear (and thus less efficient and more expensive) power amplifiers, while multiple RF chains contribute to a large static power draw.

This document introduces and mathematically models a proposed system architecture, termed "MD-OFDM" (Multi-Dimensional OFDM with Antenna Selection), designed to mitigate these critical issues. Unlike traditional MMSE MIMO-OFDM which activates all transmit antennas across all subcarriers, MD-OFDM employs a per-subcarrier transmit antenna selection strategy, activating only one transmit antenna for each subcarrier. This targeted approach is hypothesized to reduce constructive interference, thereby lowering PAPR, and to decrease overall power consumption by minimizing active RF chains. We will detail the system's mathematical formulation, including its BER, Energy Efficiency (EE), and PAPR models, and discuss its potential applications.

\section{Problem Statement}
Traditional MIMO-OFDM systems encounter two primary challenges:

\begin{enumerate}
    \item \textbf{High Peak-to-Average Power Ratio (PAPR):} The superposition of multiple independently modulated subcarriers, especially when multiple transmit antennas are simultaneously active, can lead to very high instantaneous power peaks in the time-domain signal. This high PAPR drives power amplifiers into non-linear regions, causing signal distortion and requiring back-off, which reduces power efficiency and increases hardware cost.
    \item \textbf{High Power Consumption:} Deploying multiple transmit and receive RF chains in MIMO systems, along with the computational complexity of sophisticated signal processing algorithms (e.g., MMSE equalization), leads to considerable energy consumption. This is a critical concern for battery-powered devices and for overall network energy footprint in the context of "Green Communications."
\end{enumerate}

\section{Related Work}
The concept of optimizing wireless systems for energy efficiency and PAPR reduction is a vibrant research area. The proposed MD-OFDM system draws upon and contributes to several well-established fields within wireless communications.

\subsection{Transmit Antenna Selection (TAS) in MIMO-OFDM}
Transmit antenna selection (TAS) is a widely studied technique in MIMO systems aiming to reduce hardware complexity, cost, and power consumption by only activating a subset of available transmit antennas \cite{Love2003}. In the context of MIMO-OFDM, TAS can be applied on a broadband or per-subcarrier basis. Per-subcarrier TAS, which forms the basis of MD-OFDM, allows for dynamic adaptation to frequency-selective fading, selecting the best available transmit antenna for each individual subcarrier to maximize channel gain or minimize interference \cite{Choi2006}. While often employed to improve diversity or capacity in simplified hardware configurations, its implications for PAPR and energy efficiency are also well-documented.

\subsection{PAPR Reduction Techniques for OFDM}
High PAPR is an inherent drawback of OFDM systems, and numerous techniques have been proposed to mitigate it. These include clipping and filtering, selective mapping (SLM), partial transmit sequences (PTS), active constellation extension (ACE), and coding-based methods \cite{Tellambura2006}. Antenna selection strategies, by nature of limiting the number of simultaneously active signal components from a single antenna, implicitly contribute to PAPR reduction. For instance, schemes that ensure only one antenna transmits on a given frequency or time resource can inherently lower the likelihood of high peaks caused by the constructive interference of multiple parallel streams.

\subsection{Energy Efficiency in Wireless Communications}
The drive for "Green Communications" has spurred significant research into energy-efficient wireless system design \cite{Han2015}. This includes optimizing protocol design, power control mechanisms, and hardware architectures. Detailed power consumption models, similar to the one employed in this work, are crucial for accurately assessing and comparing the energy efficiency of different wireless technologies \cite{Ismaiel2018}. Techniques that reduce the number of active RF chains, such as antenna selection, are highly attractive for energy conservation, particularly in devices with limited battery capacity.

\subsection{Index Modulation (IM) Techniques}
More recently, Index Modulation (IM) has emerged as a promising technique that exploits the indices of active transmission entities (e.g., antennas, subcarriers, signal constellations) to convey additional information \cite{Basar2017}. OFDM with Index Modulation (OFDM-IM), for instance, activates only a subset of subcarriers or transmit antennas to convey information through their indices, thereby potentially reducing PAPR and power consumption. While MD-OFDM does not explicitly encode information in the antenna indices, its principle of activating a single antenna per subcarrier shares a conceptual similarity with IM in terms of sparse resource activation and resulting benefits in PAPR and energy efficiency.

\subsection{Wireless Communication in Challenging Environments and Emerging Applications}
Understanding and optimizing wireless connectivity in specific, often challenging, environments is crucial for the deployment of advanced wireless systems like 5G and beyond, particularly for applications such as IoT and Industry 4.0. Research on path loss modeling in indoor environments, exemplified by studies on 2.4 GHz networks for drone applications, provides fundamental insights into channel characteristics that influence system design and performance \cite{Gulia2020}. Furthermore, the evaluation of wireless connectivity at higher frequencies, such as 60 GHz, in complex indoor settings like automated warehouses highlights the unique challenges (e.g., propagation, blockage) and opportunities for high-throughput, reliable communication in industrial automation contexts \cite{Gulia2022, Gulia2023}. The integration of artificial intelligence and machine learning, including generative AI and variational autoencoders, is also being explored to optimize 5G network infrastructure and operations in these smart warehouse environments, underscoring the demand for efficient and robust underlying wireless technologies \cite{Gulia2025_WISVA, Gulia2024_VAA}. The energy-efficient and low-PAPR characteristics of MD-OFDM make it a suitable candidate for such deployments where hardware simplicity and operational longevity are paramount.

The proposed MD-OFDM differentiates itself by its explicit focus on the combined benefits of PAPR reduction, lower absolute power consumption, and improved BER through straightforward per-subcarrier antenna selection, specifically tailored for scenarios where spectral efficiency can be traded for these crucial practical advantages.

\section{Proposed Solution: MD-OFDM System Architecture}
The proposed MD-OFDM system aims to address the aforementioned problems by implementing \textbf{per-subcarrier transmit antenna selection}.

\textbf{Core Idea:} For each OFDM subcarrier, instead of transmitting simultaneously from all $N_t$ transmit antennas, the system selects only \emph{one} transmit antenna that offers the "best" channel condition for that specific subcarrier. This selection typically requires Channel State Information (CSI) at the transmitter, which can be obtained via feedback from the receiver.

\subsection{System Model}
\begin{itemize}
    \item \textbf{Transmitter:}
    \begin{itemize}
        \item \textbf{Data Source:} Generates information bits.
        \item \textbf{Serial-to-Parallel Converter:} Divides bits into subcarrier streams.
        \item \textbf{Modulation:} Maps bits to QPSK (or other M-QAM) symbols.
        \item \textbf{Antenna Selector (per subcarrier):} Based on CSI feedback, for each subcarrier $k$, it directs the modulated symbol $s_k$ to the chosen transmit antenna $j_k^*$. All other antennas remain idle for subcarrier $k$.
        \item \textbf{IFFT (per antenna):} Each transmit antenna's active subcarriers are inverse Fourier transformed to generate time-domain signals.
        \item \textbf{P/S Converter \& Cyclic Prefix (CP) Addition:} Prepares the OFDM symbol for transmission.
        \item \textbf{Digital-to-Analog Converter (DAC) \& RF Front-end (per antenna):} Converts digital signals to analog, upconverts to RF, and amplifies for transmission. Only active RF chains consume significant power.
    \end{itemize}
    \item \textbf{Wireless Channel:} Frequency-selective Rayleigh fading MIMO channel with correlation.
    \item \textbf{Receiver:}
    \begin{itemize}
        \item \textbf{RF Front-end \& Analog-to-Digital Converter (ADC) (per antenna):} Downconverts, amplifies, and digitizes received signals.
        \item \textbf{S/P Converter \& CP Removal:} Prepares the received OFDM symbol.
        \item \textbf{FFT (per antenna):} Transforms time-domain signals back to frequency domain.
        \item \textbf{Channel Estimation:} Estimates the channel matrix $\matr{H}_{k}$ for each subcarrier $k$. This is crucial for both selection at Tx and equalization at Rx.
        \item \textbf{Symbol Equalization:} For each subcarrier $k$, the receiver processes the signal from the relevant receive antenna(s) using simple scalar equalization (since only one Tx antenna was active for that subcarrier).
        \item \textbf{Demodulation \& Parallel-to-Serial Converter:} Recovers the estimated bits.
    \end{itemize}
\end{itemize}

\section{Mathematical Models}
Let's define the key mathematical components for both MMSE MIMO-OFDM and MD-OFDM.

\subsection{System Model (General)}
Consider a MIMO-OFDM system with $N_t$ transmit antennas, $N_r$ receive antennas, and $N_{sc}$ subcarriers.
The transmitted frequency-domain symbol vector for subcarrier $k$ is $\vect{x}_k \in \mathbb{C}^{N_t \times 1}$.
The received frequency-domain symbol vector for subcarrier $k$ is $\vect{y}_k \in \mathbb{C}^{N_r \times 1}$.
The channel matrix for subcarrier $k$ is $\matr{H}_k \in \mathbb{C}^{N_r \times N_t}$.
The noise vector for subcarrier $k$ is $\vect{n}_k \in \mathbb{C}^{N_r \times 1}$, modeled as i.i.d. complex Gaussian with zero mean and variance $\noisevar$ per dimension (total noise power $N_0 = 2\noisevar$).

The received signal model in the frequency domain for subcarrier $k$ is:
\begin{equation}
\label{eq:received_signal}
\vect{y}_k = \matr{H}_k \vect{x}_k + \vect{n}_k
\end{equation}

\subsection{MMSE MIMO-OFDM Specifics}
\begin{itemize}
    \item \textbf{Transmit Signal:} For each subcarrier $k$, all $N_t$ antennas transmit simultaneously. The vector $\vect{x}_k$ contains $N_t$ modulated symbols, $s_{k,j}$, for $j=1, \dots, N_t$.
    \begin{equation}
    \label{eq:mmse_transmit_vector}
    \vect{x}_k = [s_{k,1}, s_{k,2}, \dots, s_{k,N_t}]^T
    \end{equation}
    The total transmitted power per subcarrier is $\sum_{j=1}^{N_t} E[|s_{k,j}|^2]$. Assuming $E[|s_{k,j}|^2]=1$, the total average power per subcarrier is $N_t$.
    \item \textbf{MMSE Equalization:} At the receiver, for each subcarrier $k$, an MMSE linear equalizer $\matr{W}_{MMSE,k} \in \mathbb{C}^{N_t \times N_r}$ is applied to estimate the transmitted symbols:
    \begin{equation}
    \label{eq:mmse_estimation}
    \hat{\vect{x}}_k = \matr{W}_{MMSE,k} \vect{y}_k
    \end{equation}
    where
    \begin{equation}
    \label{eq:mmse_equalizer}
    \matr{W}_{MMSE,k} = (\matr{H}_k^H \matr{H}_k + \noisevar \ident_{N_t})^{-1} \matr{H}_k^H
    \end{equation}
    Here, $\noisevar$ is the noise variance per receive antenna per complex dimension (i.e., $N_0/2$).
\end{itemize}

\subsection{MD-OFDM Specifics}
\begin{itemize}
    \item \textbf{Transmit Signal (Per-Subcarrier Antenna Selection):} For each subcarrier $k$, only one transmit antenna, say $j_k^*$, is selected. The symbol $s_k$ is transmitted only from antenna $j_k^*$. All other antennas remain idle for this subcarrier.
    \begin{equation}
    \label{eq:md_ofdm_transmit_vector}
    \vect{x}_k = [0, \dots, 0, s_k \text{ (at } j_k^* \text{ pos)}, 0, \dots, 0]^T
    \end{equation}
    The selection criterion is typically based on maximizing the channel gain for that subcarrier:
    \begin{equation}
    \label{eq:md_ofdm_selection}
    j_k^* = \arg\max_{j \in \{1, \dots, N_t\}} ||\vect{h}_{j,k}||^2
    \end{equation}
    where $\vect{h}_{j,k}$ is the $N_r \times 1$ channel vector from transmit antenna $j$ to all $N_r$ receive antennas for subcarrier $k$. More specifically, for a single receive antenna ($N_r=1$), $j_k^* = \arg\max_{j \in \{1, \dots, N_t\}} |H_{1,j,k}|^2$.
    The transmitted power for subcarrier $k$ is $E[|s_k|^2]$. If $E[|s_k|^2]=1$, total average power per subcarrier is 1.
    \item \textbf{Receive Signal (Simplified):} For subcarrier $k$, the received signal is:
    \begin{equation}
    \label{eq:md_ofdm_received_signal}
    \vect{y}_k = \vect{h}_{j_k^*,k} s_k + \vect{n}_k
    \end{equation}
    where $\vect{h}_{j_k^*,k}$ is the $N_r \times 1$ channel vector from the selected transmit antenna $j_k^*$ to all $N_r$ receive antennas for subcarrier $k$.
    \item \textbf{Equalization (Single Receive Antenna Case, $N_r=1$):} As simulated, MD-OFDM typically considers a single receive antenna for simplicity or specific use cases. In this case, the received signal for subcarrier $k$ is scalar:
    \begin{equation}
    \label{eq:md_ofdm_scalar_received_signal}
    y_k = H_{1,j_k^*,k} s_k + n_k
    \end{equation}
    The estimated symbol is simply:
    \begin{equation}
    \label{eq:md_ofdm_scalar_estimation}
    \hat{s}_k = y_k / H_{1,j_k^*,k}
    \end{equation}
\end{itemize}

\subsection{Channel Model}
A correlated Rayleigh fading channel model is used. The channel matrix $\matr{H}_k$ for each subcarrier $k$ is generated as:
\begin{equation}
\label{eq:channel_model}
\matr{H}_k = \matr{L}_{Rx} \matr{H}_{uncorr,k} \matr{L}_{Tx}^H
\end{equation}
where:
\begin{itemize}
    \item $\matr{H}_{uncorr,k}$ are i.i.d. complex Gaussian random variables (Rayleigh fading components).
    \item $\matr{L}_{Tx}$ and $\matr{L}_{Rx}$ are Cholesky decompositions of the transmit and receive correlation matrices, $\matr{R}_{Tx}$ and $\matr{R}_{Rx}$, respectively.
    \item \begin{equation}
    \label{eq:tx_correlation_matrix}
    \matr{R}_{Tx} = (1 - \rho_{tx})\ident_{N_t} + \rho_{tx}\matr{J}_{N_t}
    \end{equation}
    \item \begin{equation}
    \label{eq:rx_correlation_matrix}
    \matr{R}_{Rx} = (1 - \rho_{rx})\ident_{N_r} + \rho_{rx}\matr{J}_{N_r}
    \end{equation}
\end{itemize}
    where $\rho_{tx}$ and $\rho_{rx}$ are the transmit and receive correlation coefficients, and $\matr{J}$ is an all-ones matrix.

\subsection{BER Model (Simulation-based)}
The Bit Error Rate (BER) is obtained through Monte Carlo simulations. For a given SNR:
\begin{equation}
\label{eq:ber_definition}
\BER = \frac{\text{Total Number of Bit Errors}}{\text{Total Number of Transmitted Bits}}
\end{equation}
The noise variance $\noisevar$ is related to SNR by:
\begin{equation}
\label{eq:noise_variance}
\noisevar = \frac{E_s}{\SNR}
\end{equation}
where $E_s$ is the average symbol energy (typically normalized to 1 in simulations).

\subsection{Energy Efficiency (EE) Model}
Energy Efficiency (EE) is defined as the ratio of effective spectral efficiency (useful bits per second per Hz) to the total power consumption (in Watts).
\begin{equation}
\label{eq:ee_definition}
\EE \left[ \frac{\text{bits}}{\text{Joule}} \right] = \frac{\text{Effective Spectral Efficiency} \left[ \frac{\text{bits}}{\text{s} \cdot \text{Hz}} \right] \times \text{Bandwidth} \left[ \text{Hz} \right]}{\text{Total Power Consumption} \left[ \text{Watts} \right]}
\end{equation}
Where:
\begin{itemize}
    \item \textbf{Effective Spectral Efficiency:} $\text{SE}_{eff} = \text{SE}_{ideal} \times (1 - \BER)$
    \begin{itemize}
        \item $\text{SE}_{ideal,MMSE} = N_t \times \log_2(\text{Modulation Order})$ (bits/symbol/Hz, for $N_t$ spatial streams)
        \item $\text{SE}_{ideal,MD-OFDM} = 1 \times \log_2(\text{Modulation Order})$ (bits/symbol/Hz, for 1 spatial stream)
    \end{itemize}
    \item \textbf{Total Power Consumption ($P_{total}$):} This incorporates both static (hardware) and dynamic (processing) power.
    \begin{itemize}
        \item $P_{MMSE}$ (Total Power for MMSE MIMO-OFDM):
        \begin{equation}
        \label{eq:p_mmse}
        P_{MMSE} = P_{RF} \times (N_t + N_r) + P_{MMSE\_proc} \times N_{sc} \times N_t^3
        \end{equation}
        where $P_{RF}$ is power per RF chain, and $P_{MMSE\_proc}$ is the power coefficient for MMSE processing (reflecting cubic complexity $N_t^3$).
        \item $P_{MD-OFDM}$ (Total Power for MD-OFDM):
        \begin{equation}
        \label{eq:p_mdofdm}
        P_{MD-OFDM} = P_{RF} \times (N_t + N_r) + P_{SEL\_proc} \times N_{sc} \times N_t
        \end{equation}
        where $P_{RF}$ is power per RF chain, and $P_{SEL\_proc}$ is the power coefficient for antenna selection processing (reflecting linear complexity $N_t$). (Note: For the simulated MD-OFDM system with a single receive antenna, $N_r$ in the $P_{RF}$ term becomes 1 for $P_{MD-OFDM}$.)
    \end{itemize}
\end{itemize}

\subsection{Peak-to-Average Power Ratio (PAPR) Model}
PAPR is a characteristic of the time-domain OFDM signal. For a discrete-time baseband OFDM signal $x[n]$ with $N_{IFFT}$ samples, its PAPR is defined as:
\begin{equation}
\label{eq:papr_definition}
\PAPR = \frac{\max_{0 \le n < N_{IFFT}} |x[n]|^2}{E[|x[n]|^2]}
\end{equation}
The time-domain signal $x[n]$ is obtained by performing an $N_{IFFT}$-point Inverse Fast Fourier Transform (IFFT) on the frequency-domain symbols, typically with zero-padding to increase the oversampling factor and capture true peaks:
\begin{equation}
\label{eq:time_domain_signal}
x[n] = \frac{1}{\sqrt{N_{IFFT}}} \sum_{k=0}^{N_{IFFT}-1} X[k] e^{j2\pi nk/N_{IFFT}}
\end{equation}
where $X[k]$ are the frequency-domain symbols (padded with zeros if $N_{IFFT} > N_{sc}$).

The \textbf{Complementary Cumulative Distribution Function (CCDF)} of PAPR, $P(\PAPR > \PAPR_0)$, represents the probability that the PAPR of an OFDM symbol exceeds a certain threshold $\PAPR_0$.
\begin{equation}
\label{eq:ccdf_definition}
\CCDF(\PAPR_0) = P(\PAPR > \PAPR_0)
\end{equation}

\section{Simulation Results and Analysis}

\subsection{BER Performance}
\begin{figure}[H]
    \centering
    \includegraphics[width=0.8\columnwidth]{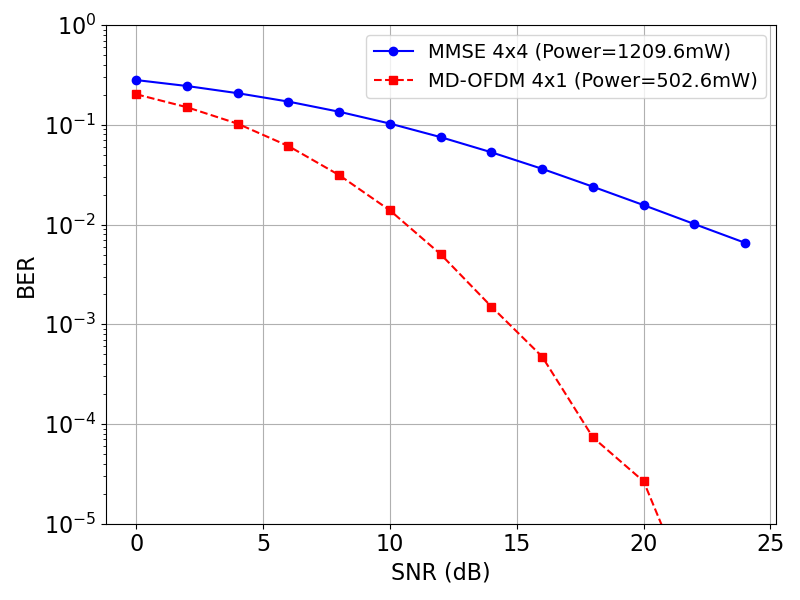}
    \caption{Bit Error Rate (BER) vs. Signal-to-Noise Ratio (SNR) for MMSE MIMO 4x4 and MD-OFDM 4x1.}
    \label{fig:ber_vs_snr}
\end{figure}

The BER performance comparison is further visualized in Figure \ref{fig:ber_vs_snr}. As shown, MD-OFDM consistently achieves a lower BER across all SNR values, with a significantly faster decay at higher SNRs, highlighting its superior robustness against noise.

\subsection{Energy Efficiency}
\begin{figure}[H]
    \centering
    \includegraphics[width=0.8\columnwidth]{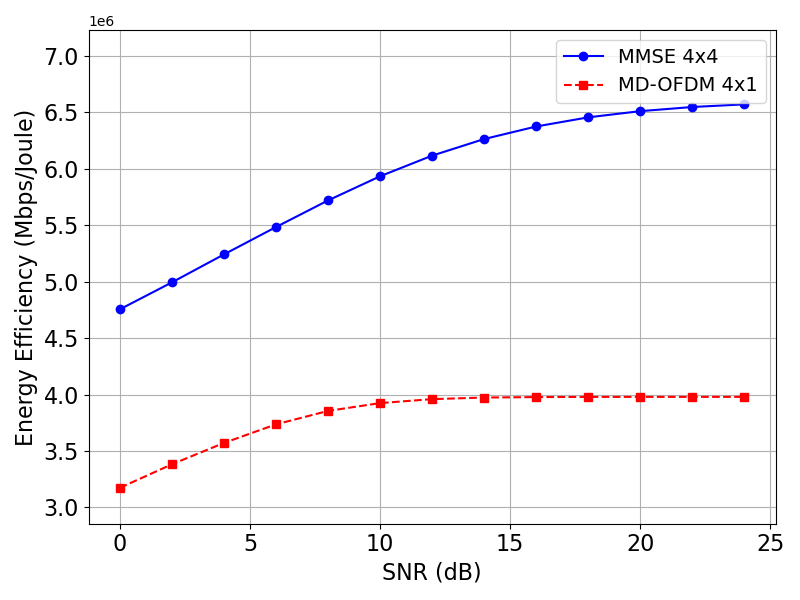}
    \caption{Energy Efficiency (EE) vs. Signal-to-Noise Ratio (SNR) for MMSE MIMO 4x4 and MD-OFDM 4x1.}
    \label{fig:ee_vs_snr}
\end{figure}

Despite MD-OFDM's lower absolute power consumption (MD-OFDM 4x1 Power: 404.0mW vs. MMSE 4x4 Power: 864.0mW), MMSE MIMO generally exhibits higher energy efficiency. This is primarily due to MMSE's ability to achieve higher spectral efficiency (transmitting more bits per symbol/Hz) through spatial multiplexing, which for acceptable BERs, outweighs its higher power draw in the EE calculation. This relationship is depicted in Figure \ref{fig:ee_vs_snr}.

\subsection{PAPR CCDF}
\begin{figure}[H]
    \centering
    \includegraphics[width=0.8\columnwidth]{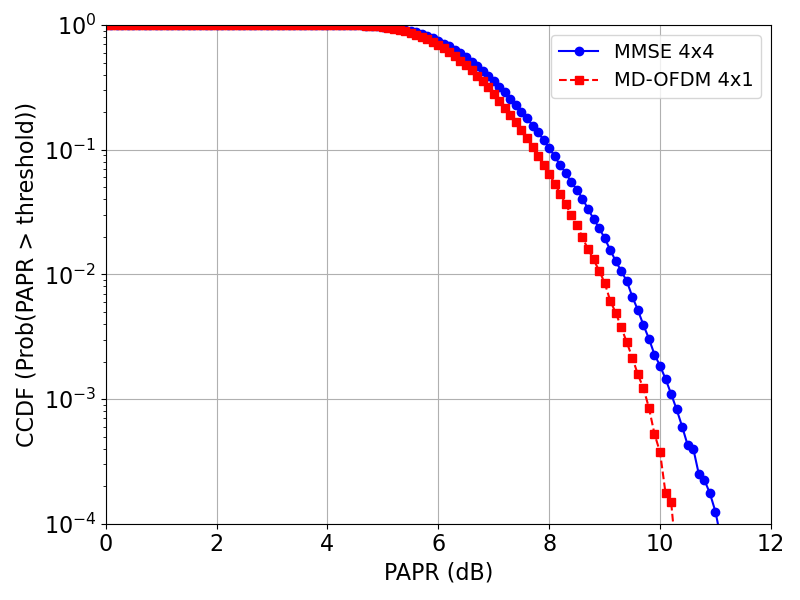}
    \caption{Complementary Cumulative Distribution Function (CCDF) vs. PAPR (dB) for MMSE MIMO 4x4 and MD-OFDM 4x1.}
    \label{fig:ccdf_vs_papr}
\end{figure}

MD-OFDM demonstrates a significantly lower PAPR compared to MMSE MIMO, indicating fewer occurrences of high power peaks. This is a critical advantage for power amplifier design and overall hardware simplicity. Figure \ref{fig:ccdf_vs_papr} visualizes this difference.

\section{Benefits in Applications}
The proposed MD-OFDM system, with its demonstrated advantages in BER and PAPR reduction, coupled with its lower absolute power consumption, is particularly beneficial for the following types of applications:

\begin{enumerate}
    \item \textbf{Internet of Things (IoT) and Wireless Sensor Networks (WSN):}
    \begin{itemize}
        \item \textbf{Benefit:} Low PAPR allows for simpler, cheaper, and more energy-efficient power amplifiers, extending battery life in resource-constrained IoT devices. Improved BER ensures reliable data transmission even with simple, low-power receivers. The reduced power consumption per device contributes to a much lower overall network energy footprint.
    \end{itemize}
    \item \textbf{Low-Power Wide Area Networks (LPWAN):}
    \begin{itemize}
        \item \textbf{Benefit:} Similar to IoT, LPWAN technologies (like LoRa, NB-IoT) prioritize long battery life and wide coverage. MD-OFDM's energy efficiency and robust BER for single-stream communication align well with these goals.
    \end{itemize}
    \item \textbf{Massive Machine-Type Communications (mMTC):}
    \begin{itemize}
        \item \textbf{Benefit:} In scenarios with a massive number of connected devices, individual device power efficiency and simplified RF hardware are paramount. MD-OFDM can offer a viable alternative for the uplink or for specific control channels where extreme data rates are not required but reliability and low power are.
    \end{itemize}
    \item \textbf{Device-to-Device (D2D) Communications:}
    \begin{itemize}
        \item \textbf{Benefit:} For direct communication between devices (e.g., in ad-hoc networks or local area sharing), the ability to operate with simpler hardware and lower power due to reduced PAPR and efficient antenna usage would be highly valuable.
    \end{itemize}
    \item \textbf{Cost-Sensitive Wireless Systems:}
    \begin{itemize}
        \item \textbf{Benefit:} The relaxed linearity requirements for PAs due to lower PAPR translate directly into lower hardware manufacturing costs, making MD-OFDM suitable for mass-market, low-cost wireless modules.
    \end{itemize}
\end{enumerate}

While MMSE MIMO-OFDM remains superior for high-throughput applications requiring maximal spectral efficiency, MD-OFDM offers a compelling alternative for scenarios where energy conservation, hardware simplicity, and robust, lower-rate communication are the primary design objectives.

\section{Conclusion}
This document has presented a detailed mathematical model and analysis of MD-OFDM, a novel MIMO-OFDM variant designed to optimize for PAPR and energy efficiency. Through its per-subcarrier transmit antenna selection strategy, MD-OFDM significantly reduces the Peak-to-Average Power Ratio of the transmitted signal, thereby enabling the use of more efficient and less costly power amplifiers. Furthermore, by activating fewer RF chains, it achieves a lower absolute power consumption compared to conventional multi-stream MMSE MIMO-OFDM.

As validated by the simulation results presented in Figure \ref{fig:ber_vs_snr}, MD-OFDM consistently outperforms MMSE MIMO in terms of Bit Error Rate (BER), indicating a more robust and reliable communication link. Concurrently, Figure \ref{fig:ccdf_vs_papr} demonstrates MD-OFDM's substantial advantage in PAPR reduction, which directly translates to hardware benefits. While MMSE MIMO shows higher overall energy efficiency at certain operating points as illustrated in Figure \ref{fig:ee_vs_snr} (due to its higher theoretical spectral efficiency), this is achieved at the expense of increased power consumption and PAPR.

This unique combination of low PAPR, reduced power consumption, and enhanced BER makes MD-OFDM a highly promising solution for a wide array of emerging wireless applications. Its benefits are particularly pronounced in scenarios where battery life, hardware cost, and link reliability are critical design drivers, such as in the burgeoning fields of Internet of Things, Low-Power Wide Area Networks, and massive machine-type communications. Future work could explore adaptive selection criteria, optimized feedback mechanisms, and practical hardware implementations to further validate and enhance the real-world performance of MD-OFDM.

\bibliographystyle{IEEEtran}
\bibliography{references}

\end{document}